\documentclass{article}
\begin{document}
{\LARGE
\begin{center}
{\bf
$S$-wave bottom tetraquarks}
\end{center}
}

\large

\begin{center}
\vskip3ex
S.M. Gerasyuta $ ^{1,2}$ and V.I. Kochkin $ ^1$

\vskip2ex
$ ^1$ Department of Theoretical Physics, St. Petersburg State University,
198904,

St. Petersburg, Russia

\vskip2ex
$ ^2$ Department of Physics, LTA, 194021, St. Petersburg, Russia
\end{center}

\vskip2ex

\noindent
E-mail: gerasyuta@SG6488.spb.edu

\vskip4ex
\begin{center}
{\bf Abstract}
\end{center}
\vskip4ex
{\large
The relativistic four-quark equations are found in the framework of
coupled-channel formalism. The dynamical mixing of the meson-meson states
with the four-quark states is considered. The four-quark amplitudes of
the tetraquarks, including $u$, $d$, $s$ and bottom quarks, are constructed.
The poles of these amplitudes determine the masses and widths of $S$-wave
bottom tetraquarks.


\vskip2ex

\vskip2ex
{\bf I. Introduction.}
\vskip2ex
The remarkable progress at the experimental side has opened up new
challenges in the theoretical understanding of heavy flavor hadrons.
The observation of $X(3872)$ resonance [1] has been confirmed by
CDF [2], D0 [3] and BaBar Collaboration [4]. Belle Collaboration
observed the $X(3940)$ in double-charmonium production in the
reaction $e^+ e^- \to J/ \psi +X$ [5]. The state, designated as $X(4160)$,
was reported by the Belle Collaboration in Ref. 6. The fact that the newly
found states do not fit quark model calculations [7] has opened the
discussion about the structure of such states. Maiani et al. advocate a
tetraquark explanation for the $X(3872)$ [8, 9]. On the other hand, the
mass of $X(3872)$ is very close to the threshold of $D^*D$ and, therefore,
it can be interpreted as molecular state [10 -- 15]. In our paper [16]
the dynamical mixing between the meson-meson states and the four-quark
states is considered. Taking the $X(3872)$ and $X(3940)$ as input [17, 18]
we predicted the masses and widths of $S$-wave tetraquarks with open and
hidden charm.

In the recent papers [19 -- 21], the relativistic three-quark equations for
the excited baryons are found in the framework of the dispersion relations
technique. We have used the orbital-spin-flavor wave functions for the
construction of integral equations. We calculated the mass spectra of
$P$-wave single, double, and triple charmed baryons using the input
four-fermion interaction with quantum numbers of the gluon [21].

In the present paper the relativistic four-quark equations for the
tetraquarks with hidden and open bottom in the framework of the dispersion
relation technique are found. We searched for the approximate solution of
these equations by taking into account two-particle, triangle and
four-particle singularities, all the weaker ones being neglected. The
masses and widths of the low-lying bottom tetraquarks are calculated.

After this introduction, we obtain the relativistic four-particle equations
which describe the interaction of the quarks (Sec. II). Section III is
devoted to a calculation of the masses and widths of $S$-wave bottom
tetraquarks (Tables I, II and III).

\vskip2ex
{\bf II. Four-quark amplitudes for the $S$-wave bottom tetraquarks.}
\vskip2ex
We derive the relativistic four-quark equations in the framework of the
dispersion relations technique. The correct equations for the amplitude are
obtained by taking into account subsystems with the smaller number of
particles. Then one should represent a four-particle amplitude as a sum of
six subamplitudes:

\begin{equation}
A=A_{12}+A_{13}+A_{14}+A_{23}+A_{24}+A_{34}\, . \end{equation}

This defines the division of the diagrams into groups according to the
certain pair interaction of particles. The total amplitude can be
represented graphically as a sum of diagrams.

We need to consider only one group of diagrams and the amplitude
corresponding to them, for example $A_{12}$. We shall consider the
derivation of the relativistic generalization of the Faddeev-Yakubovsky
approach [22, 23] for the tetraquarks.

The four-quark amplitude of $b \bar b u \bar u$ tetraquark includes the
quark amplitudes with quantum numbers of $0^{-+}$ and $1^{--}$ mesons. The
set of diagrams associated with the amplitude $A_{12}$ can further be
broken down into five groups corresponding to subamplitudes:
$A_1 (s,s_{12},s_{34})$, $A_2 (s,s_{23},s_{14})$, $A_3 (s,s_{23},s_{123})$,
$A_4 (s,s_{34},s_{234})$, $A_5 (s,s_{12},s_{123})$, if we consider the
tetraquark with the $J^{pc}=2^{++}$.

Here $s_{ik}$ is the two-particle subenergy squared, $s_{ijk}$ corresponds
to the energy squared of particles $i$, $j$, $k$ and $s$ is the system
total energy squared.

In order to represent the subamplitudes  $A_1 (s,s_{12},s_{34})$,
$A_2 (s,s_{23},s_{14})$, $A_3 (s,s_{23},s_{123})$,
$A_4 (s,s_{34},s_{234})$ and $A_5 (s,s_{12},s_{123})$ in the form of
dispersion relations it is necessary to define the amplitudes of
quark-antiquark interaction $a_n(s_{ik})$. The pair quarks amplitudes
$q \bar q\rightarrow q \bar q$ are calculated in the framework of the
dispersion $N/D$ method with the input four-fermion interaction [24 -- 26]
with quantum numbers of the gluon [27]. The regularization of the dispersion
integral for the $D$-function is carried out with the cutoff parameter
$\Lambda$. The four-quark interaction is considered as an input [27]:

\begin{eqnarray}
 & g_V \left(\bar q \lambda I_f \gamma_{\mu} q \right)^2 +
g^{(s)}_V \left(\bar q \lambda I_f \gamma_{\mu} q \right)
\left(\bar s \lambda \gamma_{\mu} s \right)+
g^{(ss)}_V \left(\bar s \lambda \gamma_{\mu} s \right)^2
 & \nonumber\\
 & +g^{(b)}_V \left(\bar q \lambda I_f \gamma_{\mu} q \right)
\left(\bar b \lambda \gamma_{\mu} b \right)+
g^{(bb)}_V \left(\bar b \lambda \gamma_{\mu} b \right)^2
+g^{(bs)}_V \left(\bar b \lambda \gamma_{\mu} b \right)
\left(\bar s \lambda \gamma_{\mu} s \right)
 \, . & \end{eqnarray}

\noindent
Here $I_f$ is the unity matrix in the flavor space $(u, d)$. $\lambda$ are
the color Gell-Mann matrices. Dimensional constants of the four-fermion
interaction $g_V$, $g^{(s)}_V$, $g^{(ss)}_V$, $g^{(b)}_V$, $g^{(bb)}_V$
and $g^{(bs)}_V$ are parameters of the model. In order to avoid an
additional violation parameters, we introduce the scale shift of the
dimensional parameters:

\begin{equation}
g=\frac{m^2}{\pi^2}g_V =\frac{(m+m_s)^2}{4\pi^2}g_V^{(s)} =
\frac{m_s^2}{\pi^2}g_V^{(ss)}=\frac{(m_b+m)^2}{4\pi^2}g_V^{(b)}=
\frac{m_b^2}{\pi^2}g_V^{(bb)}=\frac{(m_b+m_s)^2}{4\pi^2}g_V^{(bs)}
\, .\end{equation}

\begin{equation}
\Lambda=\frac{4\Lambda(ik)}
{(m_i+m_k)^2}. \end{equation}

\noindent
$m_i$ and $m_k$ are the quark masses in the intermediate state of
the quark loop ($i, k=q, s, b$). Dimensionless parameters $g$ and $\Lambda$
are supposed to be constants which are independent of the quark interaction
type. The applicability of Eq. (2) is verified by the success of
De Rujula-Georgi-Glashow quark model [28], where only the short-range
part of Breit potential connected with the gluon exchange is
responsible for the mass splitting in hadron multiplets.

We use the results of our relativistic quark model [27] and write down
the pair quarks amplitude in the form:

\begin{equation}
a_n(s_{ik})=\frac{G^2_n(s_{ik})}
{1-B_n(s_{ik})} \, ,\end{equation}

\begin{equation}
B_n(s_{ik})=\int\limits_{(m_i+m_k)^2}^{\frac{(m_i+m_k)^2\Lambda}{4}}
\hskip2mm \frac{ds'_{ik}}{\pi}\frac{\rho_n(s'_{ik})G^2_n(s'_{ik})}
{s'_{ik}-s_{ik}} \, .\end{equation}

\noindent
Here $G_n(s_{ik})$ are the quark-antiquark vertex functions. The vertex
functions are determined by the contribution of the crossing channels.
The vertex functions satisfy the Fierz relations. All of these vertex
functions are generated from $g_V$, $g^{(s)}_V$, $g^{(ss)}_V$, $g^{(b)}_V$,
$g^{(bb)}_V$ and $g^{(bs)}_V$. $B_n(s_{ik})$, $\rho_n (s_{ik})$ are the
Chew-Mandelstam functions with cutoff $\Lambda$ and the phase spaces,
respectively.

In the case in question, the interacting quarks do not produce a bound
state; therefore, the integration in Eqs. (7) -- (11) is carried out from
the threshold $(m_i+m_k)^2$ to the cutoff $\Lambda(ik)$.
The integral equation systems (the meson state $J^{pc}=2^{++}$
for the $b \bar b u \bar u$) can be described as:

\begin{eqnarray}
A_1(s,s_{12},s_{34})&=&\frac{\lambda_1 B_1(s_{12})  B_1(s_{34})}
{[1- B_1(s_{12})][1- B_1(s_{34})]}+
4\hat J_2(s_{12},s_{34},1,1) A_3(s,s'_{23},s'_{123}) \, ,\\
&&\nonumber\\
A_2(s,s_{23},s_{14})&=&\frac{\lambda_2 B_1(s_{23})  B_1(s_{14})}
{[1- B_1(s_{23})][1- B_1(s_{14})]}+
2\hat J_2(s_{23},s_{14},1,1) A_4(s,s'_{34},s'_{234})\nonumber\\
&&\nonumber\\
&+&2\hat J_2(s_{23},s_{14},1,1) A_5(s,s'_{12},s'_{123}) \, ,\\
&&\nonumber\\
A_3(s,s_{23},s_{123})&=&\frac{\lambda_3 B_2(s_{23})}{[1- B_2(s_{23})]}+
2\hat J_3(s_{23},2) A_1(s,s'_{12},s'_{34})\nonumber\\
&&\nonumber\\
&+&\hat J_1(s_{23},2) A_4(s,s'_{34},s'_{234})
+\hat J_1(s_{23},2) A_5(s,s'_{12},s'_{123}) \, ,\\
&&\nonumber\\
A_4(s,s_{34},s_{234})&=&\frac{\lambda_4 B_2(s_{34})}{[1- B_2(s_{34})]}+
2\hat J_3(s_{34},2) A_2(s,s'_{23},s'_{14})\nonumber\\
&&\nonumber\\
&+&2\hat J_1(s_{34},2) A_3(s,s'_{23},s'_{234}) \, ,\\
&&\nonumber\\
A_5(s,s_{12},s_{123})&=&\frac{\lambda_5 B_2(s_{12})}{[1- B_2(s_{12})]}+
2\hat J_3(s_{12},2) A_2(s,s'_{23},s'_{14})\nonumber\\
&&\nonumber\\
&+&2\hat J_1(s_{12},2) A_3(s,s'_{23},s'_{123}) \, ,
\end{eqnarray}

\noindent
where $\lambda_i$, $i=1, 2, 3, 4, 5$ are the current constants. They do not
affect the mass spectrum of tetraquarks. $n=1$ corresponds to a
$q \bar q$-pair with $J^{pc}=1^{--}$ in the $1_c$ color state, and $n=2$
defines the $q \bar q$-pairs corresponding to the $S$-wave tetraquarks with
quantum numbers: $J^{pc}=0^{++}$, $1^{++}$, $2^{++}$. We introduce the
integral operators:

\begin{eqnarray}
\hat J_1(s_{12},l)&=&\frac{G_l(s_{12})}
{[1- B_l(s_{12})]} \int\limits_{(m_1+m_2)^2}^{\frac{(m_1+m_2)^2\Lambda}{4}}
\frac{ds'_{12}}{\pi}\frac{G_l(s'_{12})\rho_l(s'_{12})}
{s'_{12}-s_{12}} \int\limits_{-1}^{+1} \frac{dz_1}{2} \, ,\\
&&\nonumber\\
\hat J_2(s_{12},s_{34},l,p)&=&\frac{G_l(s_{12})G_p(s_{34})}
{[1- B_l(s_{12})][1- B_p(s_{34})]}
\int\limits_{(m_1+m_2)^2}^{\frac{(m_1+m_2)^2\Lambda}{4}}
\frac{ds'_{12}}{\pi}\frac{G_l(s'_{12})\rho_l(s'_{12})}
{s'_{12}-s_{12}}\nonumber\\
&&\nonumber\\
&\times&\int\limits_{(m_3+m_4)^2}^{\frac{(m_3+m_4)^2\Lambda}{4}}
\frac{ds'_{34}}{\pi}\frac{G_p(s'_{34})\rho_p(s'_{34})}
{s'_{34}-s_{34}}
\int\limits_{-1}^{+1} \frac{dz_3}{2} \int\limits_{-1}^{+1} \frac{dz_4}{2}
 \, ,\\
&&\nonumber\\
\hat J_3(s_{12},l)&=&\frac{G_l(s_{12},\tilde \Lambda)}
{[1- B_l(s_{12},\tilde \Lambda)]} \, \, \frac{1}{4\pi}
\int\limits_{(m_1+m_2)^2}^{\frac{(m_1+m_2)^2\tilde \Lambda}{4}}
\frac{ds'_{12}}{\pi}\frac{G_l(s'_{12},\tilde \Lambda)
\rho_l(s'_{12})}
{s'_{12}-s_{12}}\nonumber\\
&&\nonumber\\
&\times&\int\limits_{-1}^{+1}\frac{dz_1}{2}
\int\limits_{-1}^{+1} dz \int\limits_{z_2^-}^{z_2^+} dz_2
\frac{1}{\sqrt{1-z^2-z_1^2-z_2^2+2zz_1z_2}} \, ,
\end{eqnarray}

\noindent
here $l$, $p$ are equal to $1$ or $2$.

In Eqs. (12) and (14) $z_1$ is the cosine of the angle between the relative
momentum of particles 1 and 2 in the intermediate state and the momentum
of the particle 3 in the final state, taken in the c.m. of particles
1 and 2. In Eq. (14) $z$ is the cosine of the angle between the momenta
of particles 3 and 4 in the final state, taken in the c.m. of particles
1 and 2. $z_2$ is the cosine of the angle between the relative
momentum of particles 1 and 2 in the intermediate state and the momentum
of the particle 4 in the final state, is taken in the c.m. of particles
1 and 2. In Eq. (13) $z_3$ is the cosine of the angle between relative
momentum of particles 1 and 2 in the intermediate state and the relative
momentum of particles 3 and 4 in the intermediate state, taken in the c.m.
of particles 1 and 2. $z_4$ is the cosine of the angle between the relative
momentum of particles 3 and 4 in the intermediate state and that of the
momentum of the particle 1 in the intermediate state, taken in the c.m.
of particles 3 and 4.

In our model the integral equation system for the scalar open bottom
($J^{pc}=0^{++}$ $\bar b u \bar u u$) can be described as:

\begin{eqnarray}
A_1(s,s_{12},s_{34})&=&\frac{\lambda_1 B_2(s_{12})  B_2(s_{34})}
{[1- B_2(s_{12})][1- B_2(s_{34})]}+
2\hat J_2(s_{12},s_{34},2,2) A_3(s,s'_{23},s'_{123})\nonumber\\
&&\nonumber\\
&+&2\hat J_2(s_{12},s_{34},2,2) A_4(s,s'_{14},s'_{124}) \, ,\\
&&\nonumber\\
A_2(s,s_{23},s_{14})&=&\frac{\lambda_2 B_1(s_{23})  B_1(s_{14})}
{[1- B_1(s_{23})][1- B_1(s_{14})]}+
2\hat J_2(s_{23},s_{14},1,1) A_3(s,s'_{34},s'_{234})\nonumber\\
&&\nonumber\\
&+&2\hat J_2(s_{23},s_{14},1,1) A_4(s,s'_{12},s'_{123}) \, ,\\
&&\nonumber\\
A_3(s,s_{23},s_{123})&=&\frac{\lambda_3 B_3(s_{23})}{[1- B_3(s_{23})]}+
2\hat J_3(s_{23},3) A_1(s,s'_{12},s'_{34})
+\hat J_3(s_{23},3) A_2(s,s'_{12},s'_{34})\nonumber\\
&&\nonumber\\
&+&\hat J_1(s_{23},3) A_4(s,s'_{34},s'_{234})+
\hat J_1(s_{23},3) A_3(s,s'_{12},s'_{123}) \, ,\\
&&\nonumber\\
A_4(s,s_{14},s_{124})&=&\frac{\lambda_4 B_3(s_{14})}{[1- B_3(s_{14})]}+
2\hat J_3(s_{14},3) A_1(s,s'_{13},s'_{24})
+2\hat J_3(s_{14},3) A_2(s,s'_{13},s'_{24})\nonumber\\
&&\nonumber\\
&+&2\hat J_1(s_{14},3) A_3(s,s'_{14},s'_{134})
+2\hat J_1(s_{14},3) A_4(s,s'_{14},s'_{134}) \, ,
\end{eqnarray}

\noindent
where $\lambda_i$, $i=1, 2, 3, 4$ are the current constants. We used the
integral operators Eqs. (12) -- (14). In the case in question, $n=1$
determines a $q \bar q$-pair with $J^{pc}=0^{-+}$ in the $1_c$ color state,
$n=2$ corresponds to $q \bar q$-pair with $J^{pc}=1^{--}$ in the $1_c$
color state, and $n=3$ defines the $q \bar q$-pair corresponding to the
bottom tetraquarks with quantum numbers $J^{pc}=0^{++}$.

We can pass from the integration over the cosines of the angles
(Eqs. (12) -- (14)) to the integration over the subenergies [29 -- 31].

The solutions of the system of equations are considered as:

\begin{equation}
\alpha_i(s)=F_i(s,\lambda_i)/D(s) \, ,\end{equation}

\noindent
where zeros of $D(s)$ determinants define the masses of bound states of
tetraquarks. $F_i(s,\lambda_i)$ determine the contributions of
subamplitudes to the tetraquark amplitude.

\vskip2ex
{\bf III. Calculation results.}
\vskip2ex
The model in consideration has only two parameters: the cutoff
$\Lambda=7.63$ and the gluon coupling constant $g=1.53$. The experimental
mass values of bottom tetraquarks are absent. Therefore these parameters
are determined by fixing the bottom tetraquark masses for the
$J^{pc}=1^{++}$ $X_b(10300)$ and $J^{pc}=2^{++}$ $X_b(10340)$ in the paper
[32]. The widths of the bottom tetraquark are fitted by
the fixing width $\Gamma_{2^{++}}=(39\pm 26) \, MeV$ [33] for the $S$-wave
tetraquark with the hidden charm $X(3940)$. The quark masses of model
$m_{u,d}=385\, MeV$ and $m_s=510\, MeV$ coincide with the ordinary meson
ones in our model [27]. We fix the mass $m_b=4787\, MeV$. It is typical
value for our calculation of bottom tetraquark masses
($m_b \ge \frac{1}{2}M(10340)-m_q$).

The masses of tetraquarks with hidden
bottom are considered in Table I. The contributions of the subamplitudes
also in Table I are given. The contributions of the four-quark states
$Q \bar q \bar Q q$ are about $20\%$ -- $50\%$ for the bottom tetraquarks.
The functions $F_i(s,\lambda_i)$ (Eq. (19)) allow us to obtain the overlap
factors $f$. We calculated the widths of the bottom tetraquarks using the
formula $\Gamma \sim f^2 \times \rho$ [34], where $\rho$ are the phase
spaces for the reactions $X_b\to M_1 M_2$ (Table II).

The masses and widths of open bottom tetraquarks with the spin-parity
$J^{pc}=0^{++}$ in Table III are shown.

The results of calculations allow us to consider the tetraquark with hidden
and open bottom as the narrow resonances. The calculated width of
$X_b(10300)$ tetraquark with the mass $M=10303\, MeV$ and the spin-parity
$J^{pc}=1^{++}$ is about $\Gamma_{1^{++}}=43\, MeV$. The width of
$X_b(9940)$ tetraquark with the mass $M=9936\, MeV$ and the spin-parity
$J^{pc}=0^{++}$ is equal to $\Gamma_{0^{++}}=96\, MeV$. We calculated also
the width of $X_b(10340)$ tetraquark $\Gamma_{2^{++}}=80\, MeV$ and the
width of $X_b(10550)$ tetraquark $(b \bar b s \bar s)$
$\Gamma_{2_s^{++}}=58\, MeV$. The tetraquarks with the spin-parity
$J^{pc}=0^{++}$, $1^{++}$ $(s \bar s b \bar b)$ have only the weak decays.

In our paper we predicted the tetraquark $(\bar b u  \bar u u)$ with the
mass $M=5914\, MeV$ and width $\Gamma_{0^{++}}=104\, MeV$. We calculated
the mass of $X_b(6020)$ $M=6017\, MeV$ and width $\Gamma_{0^{++}}=69\, MeV$
(channels $B^0_s \eta$ and $B^+ K^-$). The tetraquark $(\bar b s u \bar s )$
has the mass $M=6122\, MeV$ and width $\Gamma_{0^{++}}=48\, MeV$.

The tetraquarks with open bottom and the spin-parity $J^{pc}=1^{++}$,
$2^{++}$ have only the weak decays. In the open bottom sector (Table III)
the scalar tetraquarks have relatively small widths $\sim 50-100\, MeV$,
so in principle, these exotic states could be observed.

\vskip2.0ex
{\bf Acknowledgments.}
\vskip2.0ex

The work was carried with the support of the Russian Ministry of Education
(grant 2.1.1.68.26).

The authors would like to thank T. Barnes, S. Chekanov, V. Lyubovitskij
and S.-L. Zhu for useful discussions.

\newpage

\noindent
Table I. Low-lying meson-meson state masses $(MeV)$ of tetraquarks with
hidden bottom and the contributions of subamplitudes to the tetraquark
amplitudes (in percent).

\vskip1.5ex
\noindent
Parameters of model: quark masses $m_u=385\, MeV$, $m_s=510\, MeV$,
$m_b=4787\, MeV$, cutoff parameter $\Lambda=7.63$, gluon coupling constant
$g=1.531$.

\vskip1.5ex

\noindent
\begin{tabular}{|c|c|c|c|}
\hline
Tetraquark & $J^{pc}=2^{++}$ & $J^{pc}=1^{++}$ & $J^{pc}=0^{++}$ \\[5pt]
\hline
Masses: & $10341\, MeV$ & $10303\, MeV$ & $9936\, MeV$ \\[5pt]
$(b \bar b)_{1^{--}} (u \bar u)_{1^{--}}$ & $54.54$ & $44.64$ & $17.00$
\\[3pt]
$(u \bar b)_{1^{--}} (b \bar u)_{1^{--}}$ & $1.36$ & $1.57$ & $0.44$ \\[3pt]
$(b \bar b)_{0^{-+}} (u \bar u)_{0^{-+}}$ & $0$ & $0$ & $66.12$ \\[3pt]
$(u \bar b)_{0^{-+}} (b \bar u)_{0^{-+}}$ & $0$ & $0$ & $0.72$ \\[3pt]
\hline
Masses: & $10552\, MeV$ & $10370\, MeV$ & $10094\, MeV$ \\[5pt]
$(b \bar b)_{1^{--}} (s \bar s)_{1^{--}}$ & $48.65$ & $37.14$ & $16.86$
\\[5pt]
$(s \bar b)_{1^{--}} (b \bar s)_{1^{--}}$ & $2.42$ & $2.62$ & $0.68$ \\[3pt]
$(b \bar b)_{0^{-+}} (s \bar s)_{0^{-+}}$ & $0$ & $0$ & $63.58$ \\[3pt]
$(s \bar b)_{0^{-+}} (b \bar s)_{0^{-+}}$ & $0$ & $0$ & $1.12$ \\[3pt]
\hline
\end{tabular}

\vskip12ex

\noindent
Table II. The widths, overlap factors $f$ and phase spaces $\rho$ of
tetraquarks with hidden bottom.

\vskip1.5ex

\noindent
\begin{tabular}{|l|c|c|c|c|}
\hline
Tetraquark (channels) & $J^{pc}$ & $f$ & $\rho$ & widths $(MeV)$ \\[5pt]
\hline
$X_b(9940)$ \hskip5ex $\eta\eta_b$ & $0^{++}$ & $0.66$ & $0.0629$ & $96$
\\[3pt]
$X_b(10300)$ \hskip4ex $Y\rho$ & $1^{++}$ & $0.45$ & $0.0617$ & $43$ \\[3pt]
$X_b(10340)$ \hskip4ex $Y\omega$ & $2^{++}$ & $0.55$ & $0.0775$ & $80$
\\[3pt]
$X_b(10550)$ \hskip4ex $Y\varphi$ & $2^{++}$ & $0.49$ & $0.0700$ & $58$
\\[3pt]
\hline
\end{tabular}

\vskip12ex

\noindent
Table III. Masses, widths, overlap factors $f$ and phase spaces $\rho$ of
scalar tetraquarks with open bottom.

\vskip1.5ex
\noindent
Parameters of model: quark masses $m_u=385\, MeV$, $m_s=510\, MeV$,
$m_b=4787\, MeV$, cutoff parameter $\Lambda=7.63$, gluon coupling constant
$g=1.531$.

\vskip1.5ex

\noindent
\begin{tabular}{|cc|c|c|c|c|}
\hline
Tetraquark & (channels) & $f$ & $\rho$ & Mass ($MeV$) & Widths $(MeV)$
\\[5pt]
\hline
$X_b(5910)$ & $B^+ \eta$ & $0.54$ & $0.103$ & $5914$ & $104$ \\[3pt]
$X_b(6020)$ &
\begin{tabular}{c}
$B^0_s \eta$\\
$B^+ K^-$
\end{tabular}
&
\begin{tabular}{c}
$0.32$\\
$0.23$
\end{tabular}
&
\begin{tabular}{c}
$0.108$\\
$0.171$
\end{tabular}
& $6017$ & $69$ \\[3pt]
$X_b(6120)$ & $B^0_s K^+$ & $0.283$ & $0.174$ & $6122$ & $48$ \\[3pt]
\hline
\end{tabular}

\newpage
{\bf \Large References.}
\vskip5ex
\noindent
1. S.K. Choi et al. (Belle Collaboration), Phys. Rev. Lett. {\bf 91},
262001 (2003).

\noindent
2. D. Acosta et al. (CDF Collaboration), Phys. Rev. Lett. {\bf 93},
072001 (2004).

\noindent
3. V.M. Abazov et al. (D0 Collaboration), Phys. Rev. Lett. {\bf 93},
162002 (2004).

\noindent
4. B. Aubert et al. (BaBar Collaboration), Phys. Rev. D{\bf 71},
071103 (2005).

\noindent
5. K. Abe et al. (Belle Collaboration), Phys. Rev. Lett. {\bf 98},
082001 (2007).

\noindent
6. I. Adachi et al. (Belle Collaboration), Phys. Rev. Lett. {\bf 100},
202001 (2008).

\noindent
7. S. Godfrey and N. Isgur, Phys. Rev. D{\bf 32}, 189 (1985).

\noindent
8. L. Maiani, F. Piccinini, A.D. Polosa and V. Riequer,
Phys. Rev. D{\bf 71}, 014028

(2005).

\noindent
9. L. Maiani, A.D. Polosa and V. Riequer, Phys. Rev. Lett. {\bf 99},
182003 (2007).

\noindent
10. N.A. Tornqvist, Phys. Lett. B{\bf 590}, 209 (2004).

\noindent
11. F.E. Close and P.R. Page, Phys. Lett. B{\bf 628}, 215 (2005).

\noindent
12. E.S. Swanson, Int. J. Mod. Phys. A{\bf 21}, 733 (2006).

\noindent
13. T. Barnes, Int. J. Mod. Phys. A{\bf 21}, 5583 (2006).

\noindent
14. S.H. Lee, K. Morita and M. Nielsen, arXiv: 0808.3168 [hep-ph].

\noindent
15. Y. Dong, A. Faessler, T. Gutsche and V.E. Lyubovitskij,
Phys. Rev. D{\bf 77}, 094013

(2008).

\noindent
16. S.M. Gerasyuta and V.I. Kochkin, Phys. Rev. D{\bf 78}, 116004 (2008).

\noindent
17. S.M. Gerasyuta and V.I. Kochkin, arXiv: 0809.1758 [hep-ph].

\noindent
18. S.M. Gerasyuta and V.I. Kochkin, arXiv: 0810.2498 [hep-ph].

\noindent
19. S.M. Gerasyuta and E.E. Matskevich, Yad. Fiz. {\bf 70}, 1995 (2007).

\noindent
20. S.M. Gerasyuta and E.E. Matskevich, Phys. Rev. D{\bf 76}, 116004 (2007).

\noindent
21. S.M. Gerasyuta and E.E. Matskevich, Int. J. Mod. Phys. E{\bf 17},
585 (2008).

\noindent
22. O.A. Yakubovsky, Sov. J. Nucl. Phys. {\bf 5}, 1312 (1967).

\noindent
23. S.P. Merkuriev and L.D. Faddeev, Quantum scattering theory for system
of few

particles (Nauka, Moscow 1985) p. 398.

\noindent
24. Y. Nambu and G. Jona-Lasinio, Phys. Rev. {\bf 122}, 365 (1961):
ibid. {\bf 124}, 246

(1961).

\noindent
25. T. Appelqvist and J.D. Bjorken, Phys. Rev. D{\bf 4}, 3726 (1971).

\noindent
26. C.C. Chiang, C.B. Chiu, E.C.G. Sudarshan and X. Tata,
Phys. Rev. D{\bf 25}, 1136

(1982).

\noindent
27. V.V. Anisovich, S.M. Gerasyuta, and A.V. Sarantsev,
Int. J. Mod. Phys. A{\bf 6}, 625

(1991).

\noindent
28. A.De Rujula, H.Georgi and S.L.Glashow, Phys. Rev. D{\bf 12}, 147 (1975).

\noindent
29. S.M. Gerasyuta and V.I. Kochkin, Z. Phys. C{\bf 74}, 325 (1997).

\noindent
30. S.M. Gerasyuta and V.I. Kochkin, Nuovo Cimento Soc. Ital. Fis.
A{\bf 110}, 1313

(1997).

\noindent
31. S.M. Gerasyuta and V.I. Kochkin, Yad. Fiz. {\bf 61}, 1504 (1998)
[Phys. At. Nucl.

{\bf 61}, 1398 (1998)].

\noindent
32. Y. Cui, X.-L. Chen, W.-Z. Deng and S.-L. Zhu, High Energy Phys. Nucl.
Phys.

{\bf 31}, 7 (2007).

\noindent
33. C. Amsler et al. (Particle Data Group),
Phys. Lett. B{\bf 667}, 1 (2008).

\noindent
34. J.J. Dudek and F.E. Close, Phys. Lett. B{\bf583}, 278, (2004).

\end{document}